\documentclass[aps,prl,twocolumn]{revtex4-2}

\usepackage{graphicx}
\usepackage{colortbl}

\definecolor{darkgray}{rgb}{0.66, 0.66, 0.66}
\begin{document}

\title{Two-proton correlations in the decay of $^{48}$Ni and $^{45}$Fe}
\author{A. Ortega Moral\textsuperscript{1,2}, S.M. Wang\textsuperscript{3,4},
J. Giovinazzo\textsuperscript{1},
T. Roger\textsuperscript{5}, B. Blank\textsuperscript{1},
P. Ascher\textsuperscript{1},  C. Borcea\textsuperscript{6}, 
L. C\'aceres\textsuperscript{5}, M. Caama\~no\textsuperscript{7}, 
F. de Oliveira Santos\textsuperscript{5}, A. de Roubin\textsuperscript{8}, 
B. Fern\'andez-Dom\'inguez\textsuperscript{7}, D. Fern\'andez-Dom\'inguez\textsuperscript{7}, 
J. Lois-Fuentes\textsuperscript{7}, M. Gerbaux\textsuperscript{1}, S. Gr\'evy\textsuperscript{1}, 
M. Hukkanen\textsuperscript{1}, A. Husson\textsuperscript{1}, O. Kamalou\textsuperscript{5}, 
T. Kurtukian-Nieto\textsuperscript{1}, J. Michaud\textsuperscript{1}, J. Pancin\textsuperscript{4}, 
J. Piot\textsuperscript{5}, M. Pomorski\textsuperscript{1}, D. Regueira-Castro\textsuperscript{7}, 
A.M. S\'anchez-Benitez\textsuperscript{9}, O. Sorlin\textsuperscript{5}, M. Stanoiu\textsuperscript{6}, 
C. Stodel\textsuperscript{4}, J.-C. Thomas\textsuperscript{5}, and  M. Vandebrouck\textsuperscript{10}}

\affiliation{\textsuperscript{1} Laboratoire de Physique des Deux Infinis de Bordeaux (LP2IB), UMR 5797, 
CNRS/IN2P3, Universit\'e de Bordeaux, Chemin du Solarium, F-33170 Gradignan, France}
\affiliation{\textsuperscript{2} Consorcio IFMIF-DONES Espa\~na, Granada, Spain}
\affiliation{\textsuperscript{3} Key Laboratory of Nuclear Physics and Ion-beam Application (MOE), Institute of Modern Physics, Fudan University, Shanghai 200433, China}
\affiliation{\textsuperscript{4} Shanghai Research Center for Theoretical Nuclear Physics, NSFC and Fudan University, Shanghai 200438, China}
\affiliation{\textsuperscript{5} Grand Acc\'el\'erateur National d'Ions Lourds (GANIL), CEA/DRF-CNRS/IN2P3, Boulevard Henri Becquerel, 14076 Caen, France}
\affiliation{\textsuperscript{6} Horia Hulubei National Institute of Physics and Nuclear Engineering (IFIN-HH), 
P.O. Box MG-6, 76900 Bucharest-Magurele, Romania}
\affiliation{\textsuperscript{7} IGFAE and Dpt. de Fisica de Part\'iculas, Univ. of Santiago de Compostela, E-15758, 15 Santiago de Compostela, Spain}
\affiliation{\textsuperscript{8} Universit\'e de Caen Normandie, ENSICAEN, CNRS/IN2P3, LPC Caen, UMR6534, F-14000 Caen, France}
\affiliation{\textsuperscript{9} Departamento de Ciencias Integradas y Centro de Estudios Avanzados en 
Fisica, Matem\'aticas y Computaci\'on, Universidad de Huelva, 21071 Huelva, Spain}
\affiliation{\textsuperscript{10}
IRFU, CEA, Universit\'e Paris-Saclay, F-91191 Gif-sur-Yvette, France
}


\date{\today}

\begin{abstract}
The main observables of the rare two-proton emission process - half-life, total energy of the decay as well as energy and angular correlations between the emitted protons - have been measured for $^{48}$Ni and $^{45}$Fe in a recent experiment performed at the GANIL/LISE3 facility. The results, together with previous experimental work, are compared for the first time with calculations performed in the recently developed Gamow Coupled-Channel (GCC) framework and to 3-body predictions from literature. 
The comparison of the $^{48}$Ni and the $^{45}$Fe two-proton angular distributions with the GCC calculations confirms the adopted strength of the proton-proton interaction for both nuclei and the predominant small-angle emission. A comparison with 3-body model angular distributions indicates the shell closure of the $f_{7/2}$ orbital for $^{48}$Ni and a substantial occupancy of the $p$-orbital for $^{45}$Fe.
Discrepancies between experimental data and theoretical predictions are found when studying the other observables: half-lives, total energy of the decay and energy correlations, showing the complexity of the description of the two-proton emission process. 
\end{abstract}

\pacs{}

\maketitle


\section*{Introduction}
Since the discovery of radioactivity in 1896, decay studies have played an important role for the determination of nuclear properties. 
Two-proton radioactivity, predicted at the beginning of the 1960's by Goldanskii \cite{goldanskii-pemission}, is the simultaneous emission of two protons from the nucleus. This decay mode occurs (from the ground state) for nuclei beyond the proton drip line with an even number of protons. In this specific case, due to the pairing energy and/or structural aspects, the emission of a single proton is forbidden or highly suppressed. This process, connecting an initial state, in which the protons are localized inside the nucleus, to the final one, in which the system is spatially diffused, but still correlated, is challenging to describe. The study of the correlations between the emitted protons provides valuable information about different aspects such as the internal structure, the proton-proton interaction and the mechanism of emission. 

Several theoretical approaches have been developed focusing on the different aspects of this decay process - structure, asymptotic region or emission dynamics - predicting the observables of two-proton radioactivity. 
Some models were able to reproduce the available experimental results. The Hybrid model \cite{brown-hybrid2019} correctly predicted the half-lives of $^{45}$Fe, $^{48}$Ni, and $^{54}$Zn and the measured proton angular distribution for $^{45}$Fe was well reproduced by the 3-body model \cite{miernik-fe45}. Then, two-proton radioactivity was established for $^{67}$Kr at RIKEN \cite{goigoux2016}, with a lifetime about 20 times shorter than expected by any of the mentioned approaches. This inconsistency inspired to revisit the theoretical models, and two new theoretical approaches have been developed by L. V. Grigorenko {\it et al.} (2017) \cite{grigorenko-2000-threebody} and S. M. Wang and W. Nazarewicz (2018) \cite{nazarewicz2018}.  The former suggests that the emission in this particular case is actually a mixture of direct two-proton emission and a sequential emission. The latter, the Gamow-Coupled-Channel (GCC) model, taking the $^{48}$Ni case as a benchmark, proposes
the deformation of $^{67}$Kr to be responsible for the small measured half-life, revealing how the structure impacts the decay properties of the system in the presence of the continuum. 

Since the first experimental evidence of two-proton emission of $^{45}$Fe in 2002 at GANIL \cite{giovinazzo2002} and GSI \cite{marek2002}, the need of experimental inputs concerning two-proton emission has motivated the development of Time Projection Chamber (TPC) detectors, such as CENBG TPC \cite{cenbg-tpc}, OTPC \cite{otpc2007} or more recently ACTAR TPC \cite{roger2018,mauss2019}, aiming to measure observables otherwise inaccessible such as the individual energy of the protons and their relative emission angle. 

An experiment was performed at GANIL/LISE3 to produce the doubly-magic two-proton emitter $^{48}$Ni to study its decay properties and shell structure. The two-proton decay is characterized using ACTAR TPC.
Another two-proton emitter, $^{45}$Fe, was also produced during the experiment, and the associated decay events have also been studied. The main observables of the two-proton decay process - half-life, energy sharing and angular correlations between the emitted protons - are compared, for both $^{48}$Ni and $^{45}$Fe, with existing 3-body model calculations \cite{grigorenko-2017} and with GCC calculations performed in this work as described hereafter.

\section{Theoretical approach}
\label{chap:theoretical}

To understand the measured half-life and elucidate the correlations between emitted protons, the Gamow Coupled-Channel (GCC) approach \cite{wangPRL2021,wangjpg2022} has been employed. This method models the two-proton emitters $^{45}$Fe and $^{48}$Ni as a spherical daughter nucleus (core) plus two valence protons. The
relative motion of each pairwise component is articulated
using Jacobi coordinates with the Berggren ensemble - 
a complex basis that encompasses bound, resonant, and
scattering states. With the 3-body asymptotic behavior properly taken into account, this framework allows for a unified treatment of the structural and decay
properties of these systems. Given the relatively long lifetimes of the two-proton emitters studied in this work, the decay width and the asymptotic correlations are quantitatively assessed using the flux current \cite{wangprc2017,wangPRL2018} and a transition matrix \cite{wanginpreparation}, respectively.
In terms of the interaction between valence protons, the finite-range Furutani-Horiuchi-Tamagaki (FHT) force \cite{Furutani1979} has been adopted, supplemented by a point Coulomb force. The core-valence nuclear potential is modeled using a Woods-Saxon (WS) form with the “universal” parameter set \cite{Cwiok1987}. To align with the experimental data on two-proton decay energies, the depth of the WS potential has been adjusted, decreasing the $s$-channel strength by 7$\%$. Furthermore, the Coulomb core-proton potential is assumed to be that of a charge $Z_ce$ uniformly distributed on the spherical nuclear surface.
\section{Experimental set-up}\label{chap:setup}
The ions of interest ($^{48}$Ni and $^{45}$Fe) were produced by fragmentation of a
$75$ {\it A} MeV $^{58}$Ni beam on a 210$~\mu$m $^{\rm nat}$Ni target and selected with the LISE3 spectrometer~\cite{anne1994}, equipped with a 530 $\mu m$ achromatic beryllium degrader at the intermediate focal plane.
Some ancillary detectors were placed along the beam-line, for the identification of the fragments: three position-sensitive gaseous detectors (CATS \cite{cats1999}) for time-of-flight and position measurements, two of them located after the first two dipoles and a third one after the third dipole. Besides, three different silicon detectors were located before ACTAR TPC (S1: 140 $\mu m$, S2: 140 $\mu m$, S3: 300 $\mu m$ thick) for energy measurements. The implantation of the isotopes of interest in ACTAR TPC was adjusted using an aluminum degrader placed in the beam-line, about 1~m upstream ACTAR TPC.

The implanted nuclei were identified in mass ($A$) and charge ($Z$) on
an event-by-event basis from a multi-parameter analysis, using the cyclotron radio-frequency and beam-line detectors for time-of-flight measurements, two different energy-loss values measured by silicon detectors and the horizontal position X in one of the CATS detectors. The expected values of these parameters for the nuclei of interest, $^{48}$Ni and $^{45}$Fe, were extrapolated, due to their low production rates, from the parametrization of less exotic isotopes also implanted in the detector ($^{42,41}$Ti, $^{44,43}$V, $^{45,44,43}$Cr, $^{
46}$Mn) as in~\cite{dossat-2005}. 
In order to increase the counting rate of the most exotic species during the experiment, the  momentum slits of LISE3 were widely opened to $\pm$ 25 mm. This leads to an overlap in the time-of-flight identification parameter between neighbouring isotopes, thus making the separation of e.g. $^{45}$Fe and $^{46}$Fe difficult. A clean identification has been obtained by imposing extra contour cuts in the time-of-flight and the S1 silicon detector two-dimensional identification matrix. This identification criterium has been established by studying $\beta$-delayed proton emission of stronger produced isotopes during the experiment ($^{45}$Cr, $^{44}$Cr, $^{43}$Cr and $^{46}$Fe, $^{47}$Fe), revealing a negligible contamination of lower neutron number ({\it N}) isotopes from higher {\it N} isotopes, which is the case in the current work for $^{48}$Ni and $^{45}$Fe and their closest neighbours $^{49}$Ni and $^{46}$Fe, respectively.

\begin{figure}[!ht]
\begin{center}
\begin{minipage}{9pc}
\includegraphics[width=1.0\textwidth]{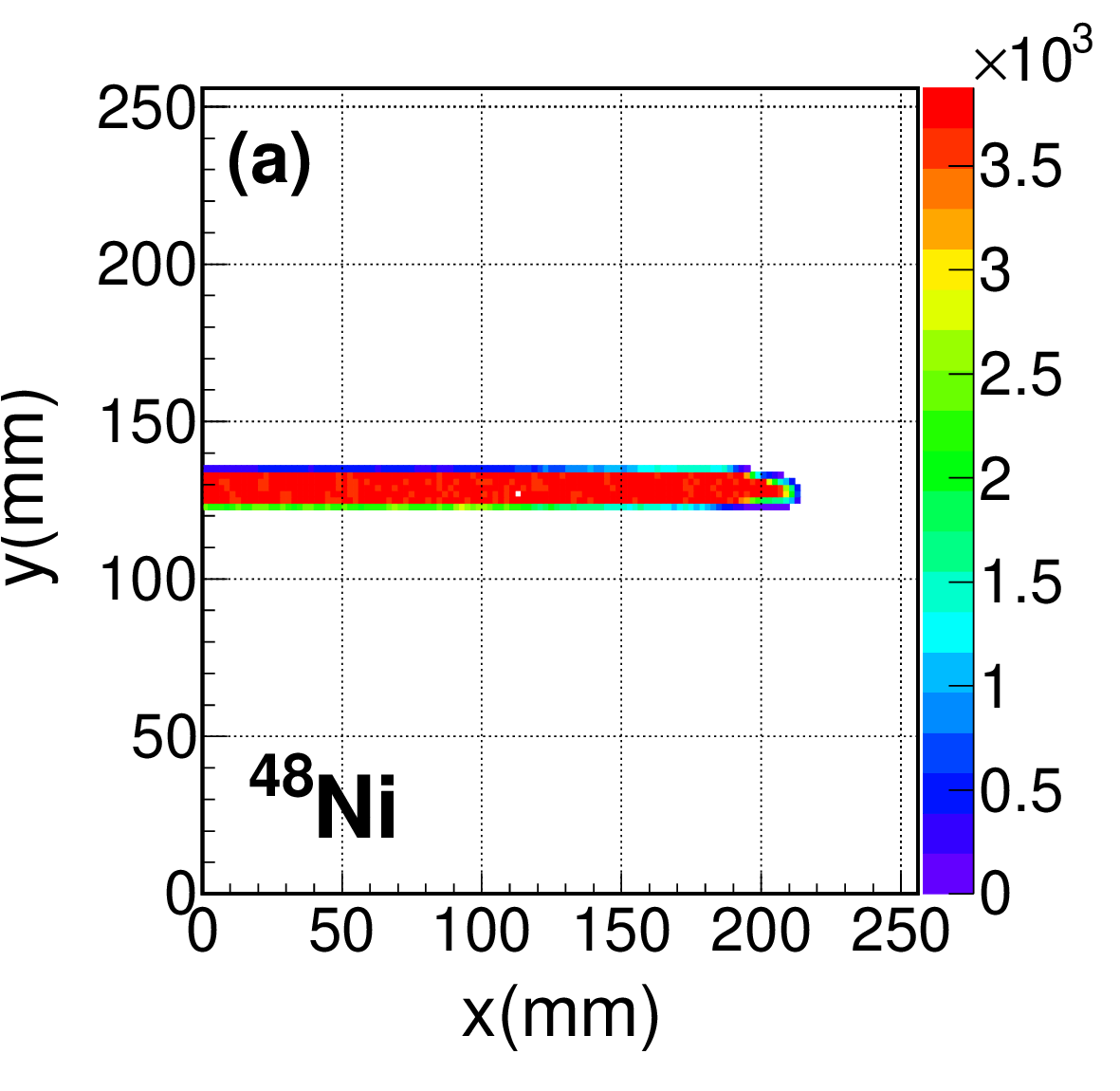}
\end{minipage}
\begin{minipage}{9pc}
\includegraphics[width=1.0\textwidth]{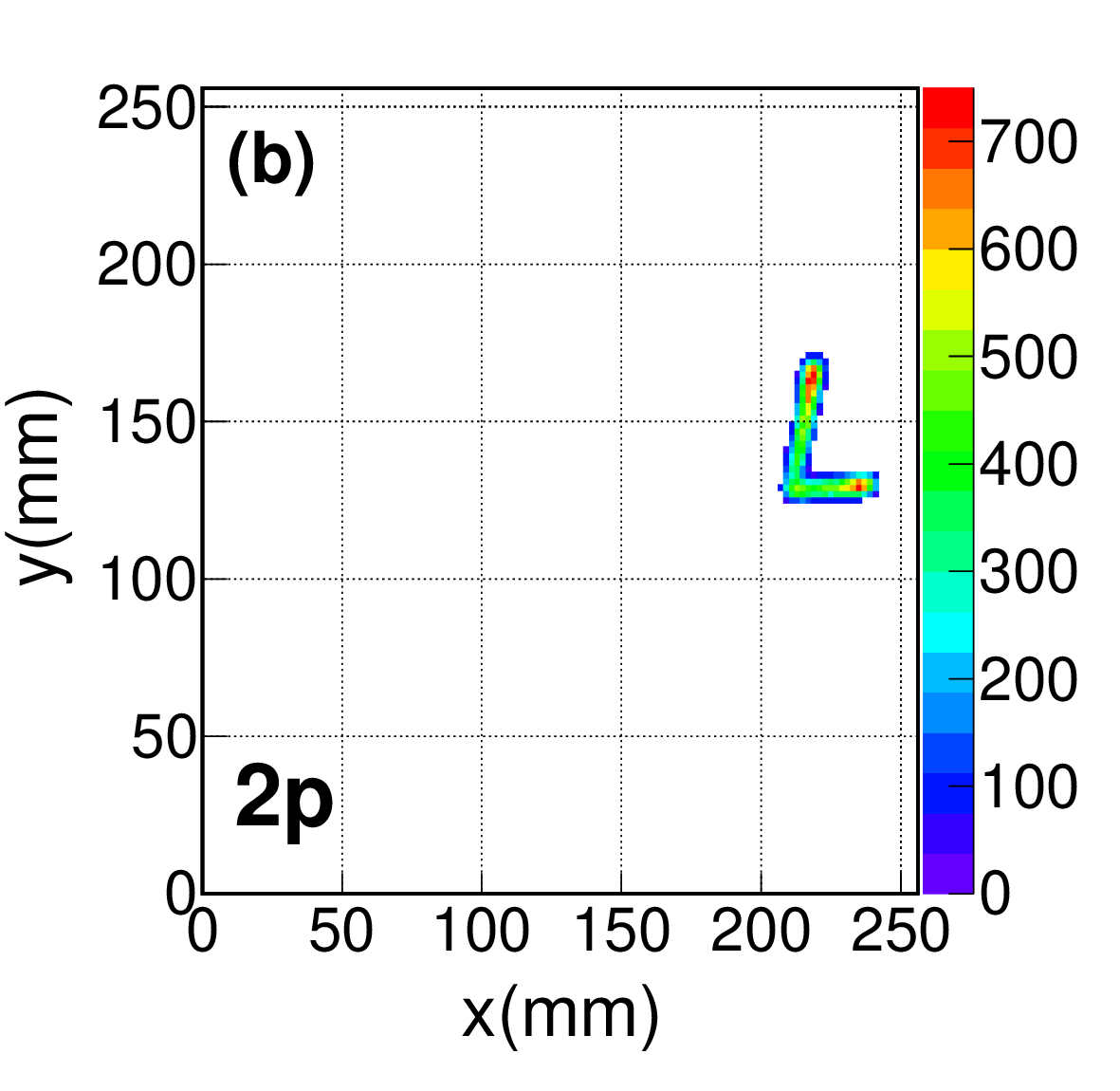}
\end{minipage}
\begin{minipage}{9pc}
\includegraphics[width=1.0\textwidth]{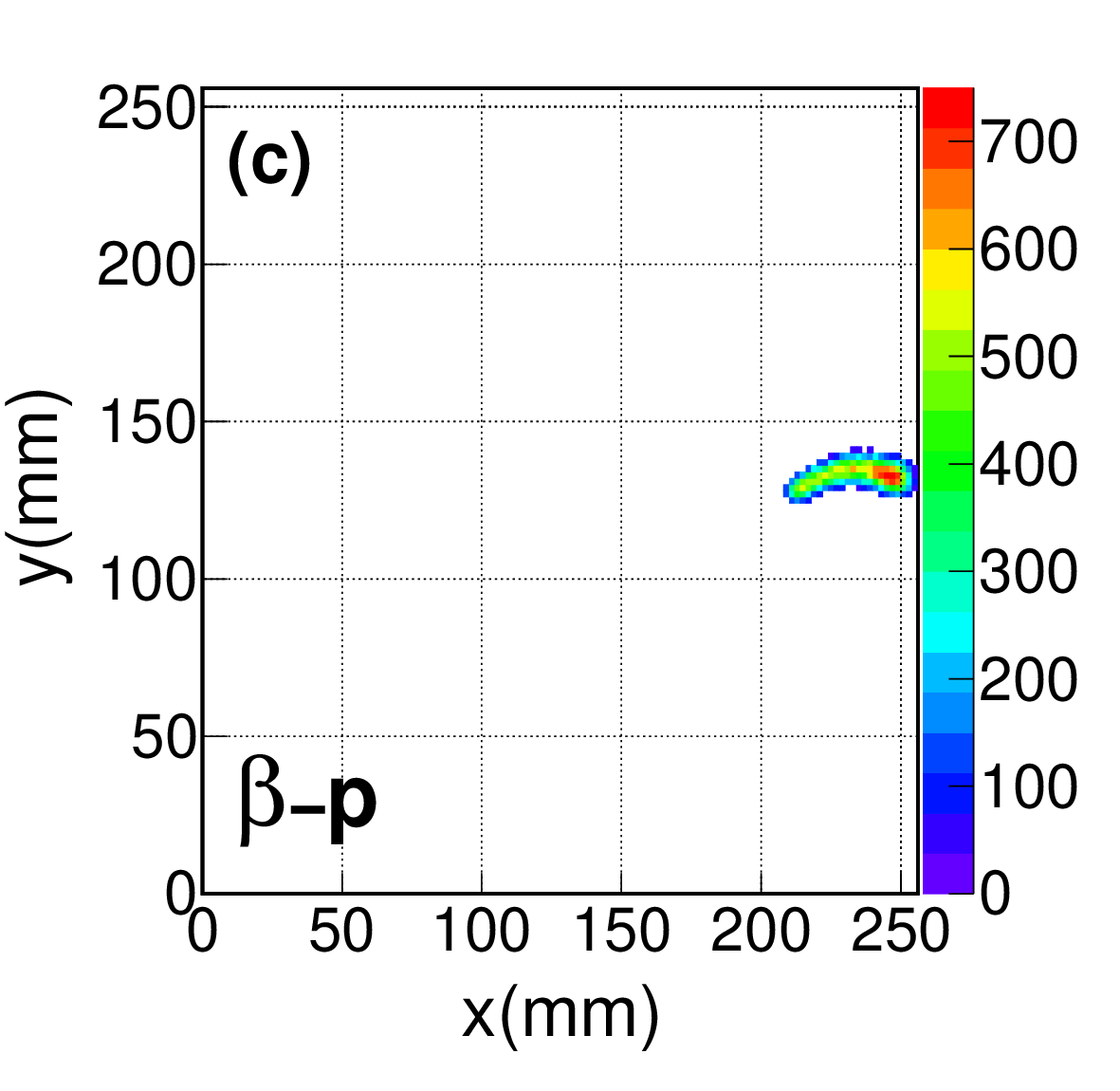}
\end{minipage}
\caption{Collected charges on the pad plane of three consecutive events recorded in ACTAR TPC: implantation of a $^{48}$Ni nucleus (upper left panel), two-proton decay (upper right panel) and proton emission from the $\beta$-$p$ decay of the daughter nucleus $^{46}$Fe (bottom panel). The time differences from the implantation event are 2.1 ms and 9.8 ms for the second and third event, respectively. \label{fig:eventexampleni48}}
\end{center}
\end{figure}

The ACTAR TPC device allows for a full 3D reconstruction of the proton tracks. 
In the current experiment, the chamber was filled with argon (90\%) and isobutane (10\%). The pressure was set to 300 mbar in the first part of the experiment and 450 mbar in the second part.
The ion implantation events are triggered by the S1 silicon detector. When implanted in the active gas volume of ACTAR TPC, the subsequent decay events are triggered by the internal trigger of the TPC readout electronics \cite{pollacco2018}. The correlation between these two different types of events is performed by imposing a time condition of 10 half-lives ($T_{1/2}$) and requiring a maximum distance between the stopping point of the implantation track and the starting point of the proton track of 8 mm.
From the 3D analysis of the proton tracks \cite{jerome2022}, the angles between the emitted protons and the individual track lengths were determined. The individual energy of the protons is obtained by conversion of the measured lengths into energy units using energy-loss tables (SRIM \cite{SRIM}), whose parameters (pressure and temperature) are optimized to reproduce well-known proton energies from decay events of $^{41}$Ti,  produced during the experiment. The total energy of the decay for two-proton emission $Q_{2p}$ is cross-checked in two different ways in this work: (i) by summing the individual proton energies determined from the track lengths and the recoil energy ($Q_{2p}(L)$); (ii) from the total charge deposit in ACTAR TPC ($Q_{2p}(C)$), also calibrated to reproduce well-known $\beta$-delayed proton energies. Typical implantation and decay events are shown in Figure~\ref{fig:eventexampleni48}.

\section{Results}
\label{chap:results}
\subsection{$^{48}$Ni}
Seven events have been identified as $^{48}$Ni during an effective measurement time of 213 h. Six of them were correctly implanted in the detection volume, from which five decay events were measured. Three of them were identified as two-proton emissions, one as $\beta$-delayed three proton emission (first observation of this decay mode for $^{48}$Ni), and one as $\beta$-delayed single proton emission. This leads to branching ratio values of  $BR(2p)=60^{+25}_{-30}\%$, $BR(\beta$-$p)=20^{+32}_{-17}\%$ and $BR(\beta$-3$p)=20^{+32}_{-17}\%$, in agreement with previous results obtained by Pomorski {\it et al.}~\cite{Pomorski-ni48} and Dossat {\it et al.}~\cite{dossat-2005}. No signal from ACTAR TPC was registered after one of the implantation events. However, a scaler module counting the ACTAR TPC events indicates that, most likely, both the decays of $^{48}$Ni and of the daughter nucleus may have happened during the dead time of the acquisition system. 
The lifetime is obtained as suggested by Schmidt {\it et al.} \cite{pocascuentas} from the average time between implantation and decay, shifting this value by the average dead time per event (0.225 ms) of the acquisition, leading to a half-life value of $T_{1/2}=$1.32$^{+1.06}_{-0.41}$$~$ms, in agreement with previous experimental values \cite{Pomorski-ni48,dossat-2005}.

\begin{figure}[!ht]
\begin{center}
\begin{minipage}{19pc}
\centerline{\includegraphics[width=1.2\textwidth]{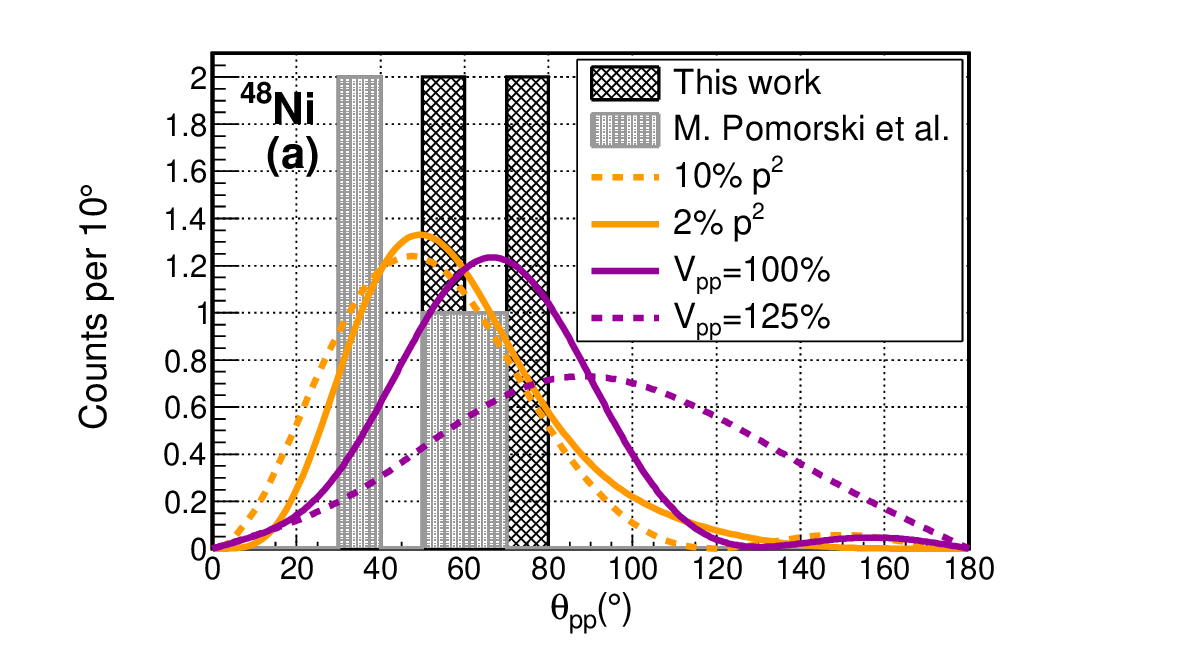}}
\end{minipage}
\begin{minipage}{19pc}
\centerline{\includegraphics[width=1.2\textwidth]{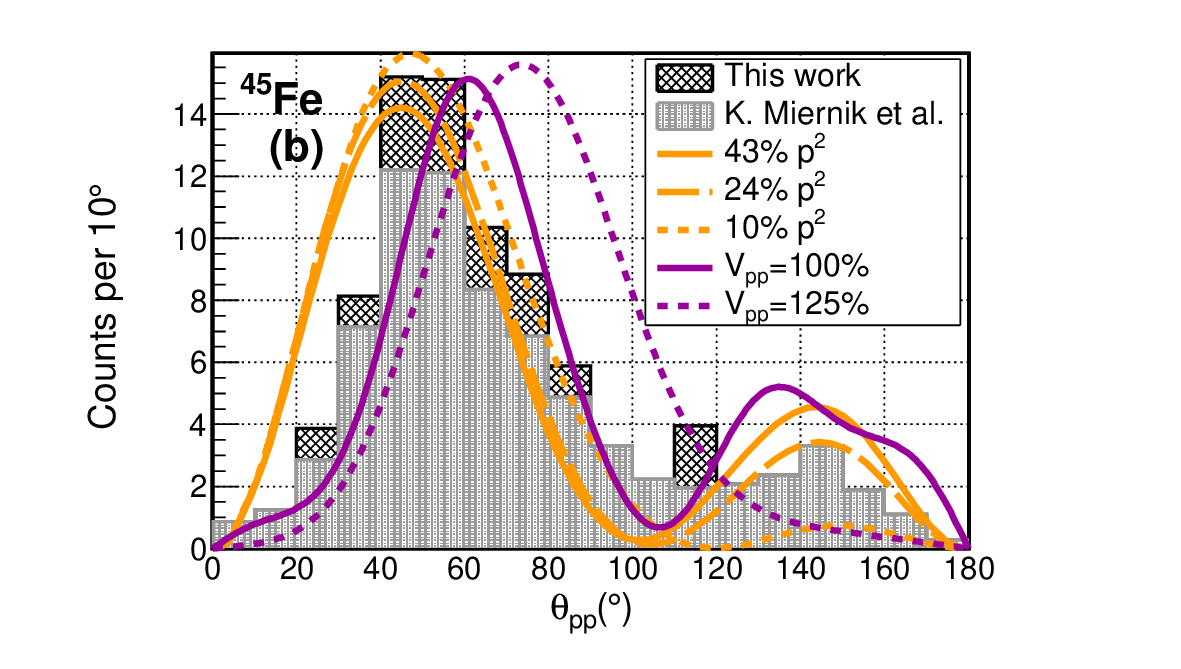}}
\end{minipage}
\caption{
Reconstructed angles between the two protons emitted from $^{48}$Ni (a) and $^{45}$Fe (b) for the current work added on top of those reconstructed in \cite{Pomorski-ni48} and \cite{miernik-fe45}, respectively. Theoretical predictions (3-body \cite{grigorenko-2000-threebody} and GCC of this work) are represented in orange and purple, respectively.
\label{fig:angulardistributions}}
\end{center}
\end{figure}

From the three two-proton emissions associated with $^{48}$Ni, the angles between the protons were measured. The uncertainties are estimated using a simulation. The values are shown in Table \ref{tabla-anglesni48} and represented in Figure \ref{fig:angulardistributions}(a) (black) added on top of previous results$~$\cite{Pomorski-ni48} (gray) for the comparison with theory. 

\begin{figure}[!ht]
\begin{center}
\begin{minipage}{19pc}
\centerline{\includegraphics[width=1.2\textwidth]{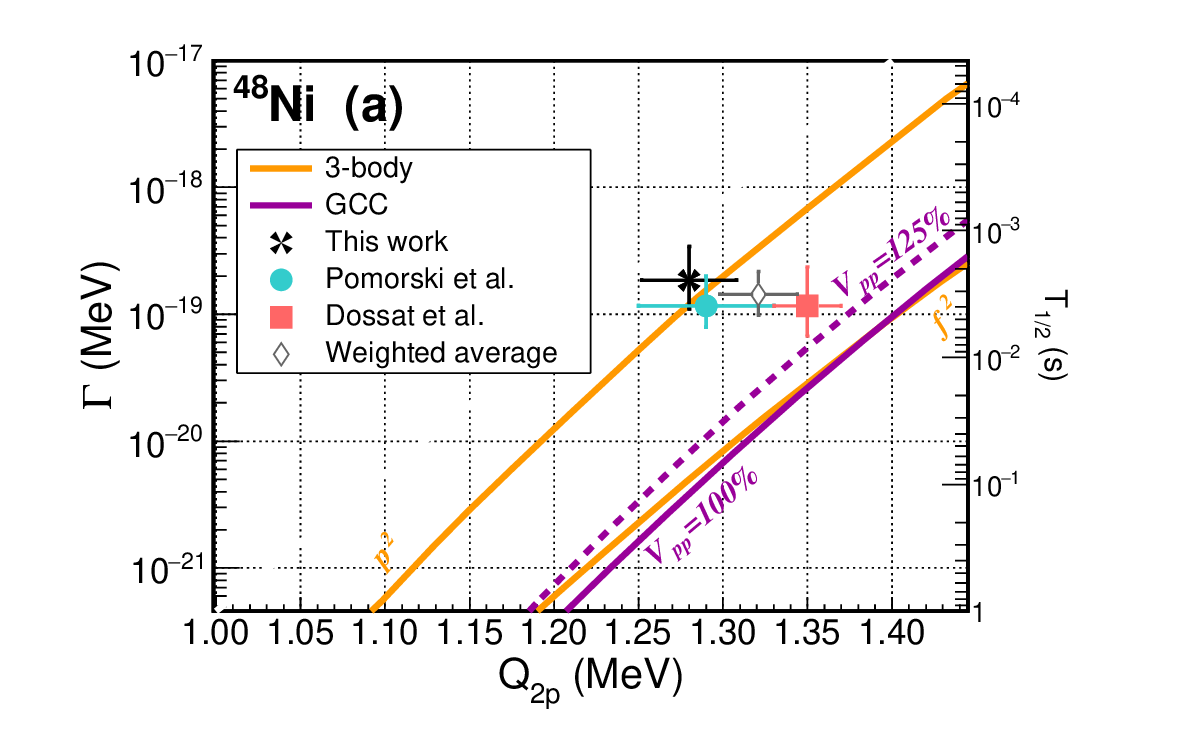}}
\end{minipage}
\begin{minipage}{19pc}
\centerline{\includegraphics[width=1.2\textwidth]{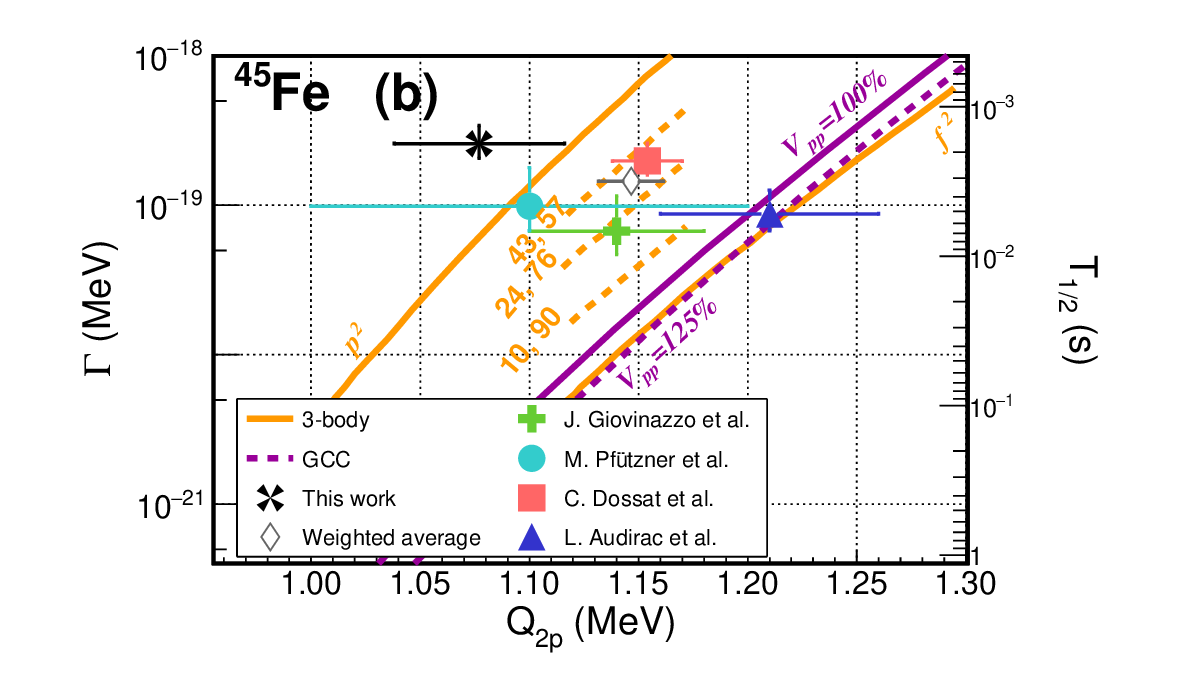}}
\end{minipage}
\caption{Experimental decay widths, partial two-proton half-lives and total decay energies measured in this work (in black) and from previous work are represented for $^{48}$Ni~(a)~\cite{dossat-2005,Pomorski-ni48}, and $^{45}$Fe (b)~\cite{dossat-2005,giovinazzo2002,marek2002,audirac-2012}, respectively. The weighted average result from all values is also represented by the open symbol. Predicted values under different assumptions from the 3-body model~\cite{grigorenko-2017} and the GCC calculations (this work) are plotted in orange and purple. 
\label{fig:half-lives}}
\end{center}
\end{figure}

The energy of individual protons and the total energy of the decay ($Q_{2p}$) were measured only for one of the two-proton events, in which both protons stayed confined in the detection volume. 
The $Q_{2p}$ value is calculated in two different ways as previously explained, yielding $Q_{2p}(L)$=1286(30) keV and $Q_{2p}(C)$=1227(90) keV. Since these values are calculated for a single event, the errors have been approximated by the resolution of the detector for both techniques. Both values are compatible and the average value, $Q_{2p}$=1280(28) keV, is established as the $Q_{2p}$ value measured in this work. Partial $T_{1/2}$ and $Q_{2p}$ values are represented together with previous experimental results \cite{Pomorski-ni48,dossat-2005} in Figure \ref{fig:half-lives}(a). The partial half-life values are calculated from measured half-lives from each experiment and the averaged branching ratio $BR(2p)=54^{+14}_{-13}\%$ computed for all experiments.
 The experimental partial half-life obtained in the present work is consistent with previous experimental values. The $Q_{2p}$ value is in agreement with a previous TPC measurement by Pomorski \textit{et al.} \cite{Pomorski-ni48}, but significantly smaller than the value measured with a silicon detector by Dossat \textit{et al.} \cite{dossat-2005} based on a single event, as can be seen in Figure~\ref{fig:half-lives}. 

\begin{table}[!h]
  \setlength\arrayrulewidth{1.1pt}
 \arrayrulecolor{darkgray}
\centering
\caption{\label{tabla-anglesni48} Reconstructed proton energies (when possible) and angles between emitted protons for three decay events of $^{48}$Ni from the present experiment. 
}
 \begin{tabular}{c|c|c|c|c}
 \hline
 \hline
   $Q_{2p}(L)$ (keV)& $Q_{2p}(C)$ (keV)& $E_{p_1}$ (keV)& $E_{p_2}$ (keV)& $\Theta_{pp}$ ($^o$)  \\
      \hline 
 
    1286(30)&1227(90) &679(16) & 533(12)& 71(10)  \\
    - &- & - &  -& 58(10)  \\
     - & - &- &  -& 72(10)  \\
\hline
    \hline
\end{tabular}
\end{table}

\subsection{Comparison with theory}
 The partial half-life and total decay energy obtained from this experiment, as well as the two-proton angular distribution for $^{48}$Ni, obtained from the present work and Pomorski {\it et al.}~\cite{Pomorski-ni48}, are compared with theoretical predictions from the 3-body model~\cite{grigorenko-2000-threebody}, GCC calculations and $Q_{2p}$ predictions by Brown, Ormand and Cole~\cite{brown-ni48,ormand-ni48,cole-ni48} in Table \ref{ni48andfe45-q2pvaluestable}. The $Q_{2p}$ value is compatible with theoretical predictions by Brown \cite{brown-ni48} and Ormand \cite{ormand-ni48}.

The angular distribution between the two protons from the decay of $^{48}$Ni (current results on top of previous results \cite{Pomorski-ni48}) is compared for the first time to theory in Figure \ref{fig:angulardistributions}(a).
Novel GCC calculations considering different strengths of the proton-proton interaction, $V_{pp}$=100\% and $V_{pp}$=125\% (see \cite{minnesotaforce}), are performed in this work, at longer distances from the core with respect to those in \cite{nazarewicz2018}. This corresponds to configurations with 92.6\% $f^2$ and 5.7\% $p^2$ contribution for $V_{pp}$=100\% as well as 86.1\% $f^2$ and 10.5\% $p^2$ contribution for $V_{pp}$=125\%. The distribution of the relative angles is also compared to 3-body model predictions in literature made for $^{45}$Fe and $^{54}$Zn, considering a low occupancy of the $p^2$ orbital, $\omega(p)$=10$\%$ and $\omega(p)$=2$\%$, respectively. According to the standard shell-model picture, $^{48}$Ni is a closed-shell nucleus, in which the $f_{7/2}$ orbital is fully occupied and the $p$-wave component is expected to be small.
The distribution of the relative angles, predicted by the 3-body and the GCC models, are plotted in Figure \ref{fig:angulardistributions}(a) in orange and purple, respectively, normalized to the total number of observed events. 
The likelihood probability $L$ and the $\chi^2$ value per degree of freedom ($\chi^2 \big/dof$) obtained when comparing these calculations with the experimental results are shown in Table \ref{tabla-chi2-ni48}. Similar agreement is found for the 3-body model and for the GCC calculations using a standard nucleon-nucleon interaction of $100\%$ $V_{pp}$, although higher statistics is certainly needed to conclude. 

\begin{table}[!h]
  \setlength\arrayrulewidth{1.1pt}
 \arrayrulecolor{darkgray}
\centering
\caption{\label{ni48andfe45-q2pvaluestable} Comparison of experimental $Q_{2p}$ values with theoretical predictions for $^{48}$Ni and $^{45}$Fe. All values are in keV. Average values from \cite{Pomorski-ni48,dossat-2005} for $^{48}$Ni  and \cite{giovinazzo2002,marek2002,dossat-2005,miernik-fe45,audirac-2012} for $^{45}$Fe. 
}
\begin{tabular}{c|c|c|c|c|c}
 \hline
 \hline
    & This   & Prev.works & Brown & Ormand & Cole  \\
    &work   & (avg) & \cite{brown-ni48} & \cite{ormand-ni48} & \cite{cole-ni48} \\
      \hline 
 $^{48}$Ni &1280(28) & 1338(24) & 1360(130)& 1290(330)&1350(60)  \\
\hline
$^{45}$Fe & 1077(39)  &1156(14) & 1154(94) &1279(181)
    &1218(49)  \\
    \hline
    \hline
\end{tabular}
\end{table}

\begin{table}[!h]
  \setlength\arrayrulewidth{1.1pt}
 \arrayrulecolor{darkgray}
 \caption{ \label{tabla-chi2-ni48} Likelihood probabilities and $\chi^2$ per degree of freedom values obtained when comparing experimental (present work and Pomorski {\it et al.}~\cite{Pomorski-ni48}) and theoretical angular distributions for $^{48}$Ni. 
}
\centering
 \begin{tabular}{c|c|c}
\hline
\hline
  Model& $L(\theta_{pp}$)&$\chi^2 \big/dof$\\  
  \hline 
    GCC $100\% V_{pp}$ & 4.7$\times10^{-5}$ &0.54 \\
    GCC $125\% V_{pp}$ & 4.8$\times10^{-7}$ &1.38\\
    3-body $\omega(p)=10\%$ &  7.5$\times10^{-5}$ &0.40\\
    3-body $\omega(p)=2\%$ &  9.8$\times10^{-5}$&0.37\\
    \hline
    \hline
\end{tabular}
\end{table}

The non-observation of large angles of emission may indicate, despite the reduced statistics, the persistence of the $Z=28$ shell closure for $^{48}$Ni (3-body model comparison) and favours the normal strength (100\%) of the nucleon-nucleon potential for $^{48}$Ni. In contrast, when comparing predicted and experimental half-life values, better agreement is found for a 100$\%$ $p^2$ configuration (3-body model) and a strong $V_{pp}$ potential (GCC model), which leads to conclusions opposite to those from the angular distribution. 
Further measurements with higher statistics and new theoretical studies may help for the understanding of these discrepancies in the interpretation of the results of two-proton emission observables. 

\subsection{$^{45}$Fe}

Seven events were identified as $^{45}$Fe and correctly implanted in ACTAR TPC. Six of the associated decays were identified as two-proton emissions and one as a $\beta$-$p$ decay. This leads to a two-proton branching ratio of $BR(2p)=86^{+12}_{-26}\%$, compatible with previous experimental values \cite{miernik-fe45}. The decay events of the neighbouring isotope $^{46}$Fe are also analysed and ten of them have been characterized as two-proton decays~\footnote{The same analysis was also performed for $^{49}$Ni, but no two-proton events were found in this case.} and are thus re-identified as $^{45}$Fe decays instead. This is due to the identification problem mentioned in the description of the experimental setup. Because $\beta$-x(p) emissions cannot be distinguished to be emitted from $^{46}$Fe or $^{45}$Fe, these 10 two-proton emission events cannot be taken into account for the branching ratio determination, since they will bias this value to a higher two-proton emission branching ratio. They have been included for the determination of the other observables: half-life, individual proton energies, total decay energy and angle between the emitted protons of $^{45}$Fe. The half-life is obtained from a decay-law fit, $T_{1/2}(fit)$=1.31(37) ms, and using the Schmidt method, $T_{1/2}(Schmidt)$=1.22$^{+0.39}_{-0.24}$$~$ms. Both values are in agreement, and the second one is established as the $T_{1/2}$ value measured in this work to avoid a small dependence from the time range considered for the fit.
This value is smaller than the averaged half-life value obtained in previous experiments \cite{giovinazzo2002,marek2002,audirac-2012,miernik-fe45} but in good agreement with the value from Dossat {\it et al.}~\cite{dossat-2005}. 
The angles between the emitted protons could be reconstructed for 15 events and are represented in Figure \ref{fig:angulardistributions}(b), on top of previous results measured in \cite{miernik-fe45}. The angle could not be reconstructed for one two-proton event, because of the small track length and the direction of emission of the protons. 

Individual proton energies could be measured for 14 of these two-proton events. As in the case of $^{48}$Ni, two different methods of $Q_{2p}$ determination are employed, leading to $Q_{2p}(L)$=1116(17) keV and $Q_{2p}(C)$=1039(17) keV. The average result, $Q_{2p}$=1077(39) keV, is established as the $Q_{2p}$ value measured in this work. This $Q_{2p}$ value is smaller than those reported in previous experiments \cite{dossat-2005,giovinazzo2002,marek2002,audirac-2012}, but compatible with some of them \cite{giovinazzo2002,marek2002} as it can be seen in Figure \ref{fig:half-lives}(b). As this is also the case for $^{48}$Ni, the TPC measurement seems to be energy-shifted with respect to the silicon detector values in the case of two-proton events. To exclude a bias coming from the analysis method and test such a hypothesis, our experimental analysis procedures have been applied to simulated events, and the correct input values of the simulation have been found.
The partial half-life has been calculated from the current $T_{1/2}$ value and the averaged branching ratio $BR(2p)=69^{+3}_{-4}\%$ of all available experimental measurements \cite{dossat-2005,audirac-2012,miernik-fe45} including the present branching ratio. This value is compared in Figure \ref{fig:half-lives}(b) with previous experimental results where both, the half-life and the $Q_{2p}$ values were measured \cite{giovinazzo2002,marek2002,dossat-2005,audirac-2012}. The partial half-lives of these experiments have also been computed with the averaged branching ratio. 

\begin{figure}[!ht]
\begin{minipage}{19pc}
\begin{center}
\centerline{\includegraphics[width=1.2\textwidth]{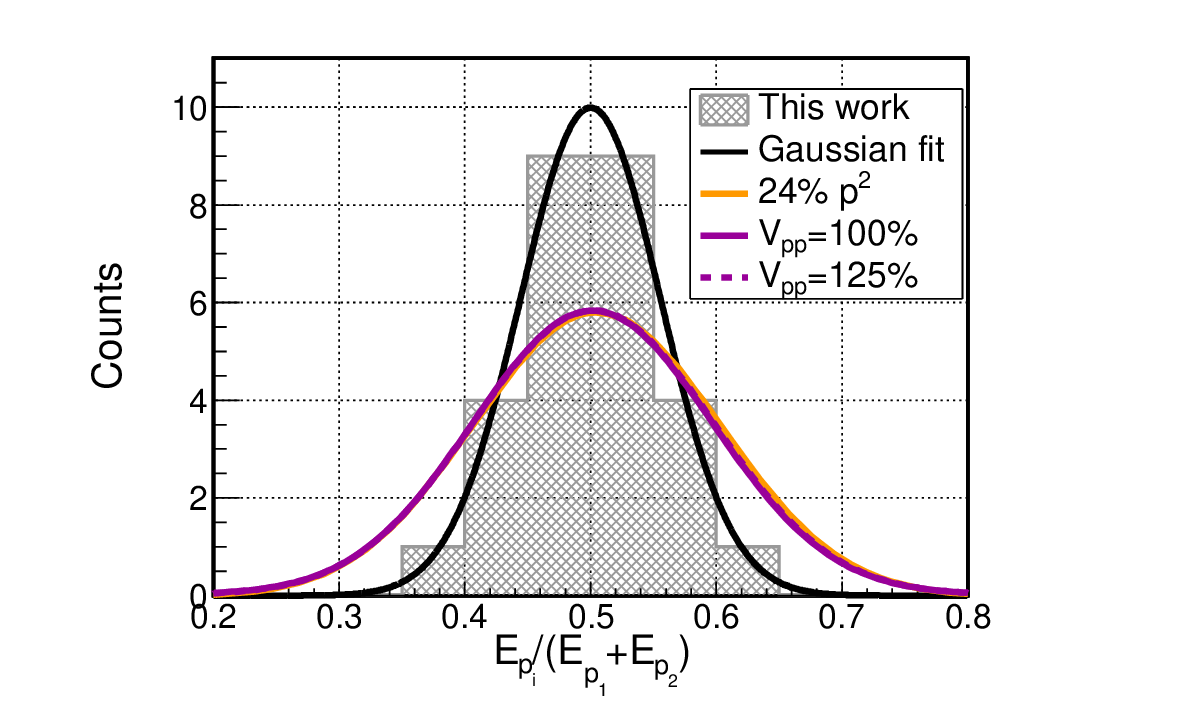}}
\caption{Distribution of individual proton energies $E_{p_{i}}$ from $^{45}$Fe normalised to the sum of both proton energies on an event-by-event basis for the present work. The orange and purple curves indicate the predictions of the 3-body model \cite{grigorenko-2000-threebody} and the GCC model \cite{gcc-2023}, respectively. The black curve is a Gaussian fit of the experimental distribution \label{fig:energy45Fe}}
\end{center}
\end{minipage}
\end{figure}

\subsection{Comparison with theory}
All the two-proton observables, partial half-life, total decay energy and two proton correlations (energy and angle) are compared with the 3-body model and GCC calculations performed for $^{
45}$Fe in this work. As in the case of $^{
48}$Ni, the calculations have been performed for different $V_{pp}$ values, leading to different weights in the orbital configuration: $V_{pp}$=100\% (82.5\% $f^2$, 13.7\% $p^2$) and $V_{pp}$=125\% (88.1\% $f^2$, 8.7\% $p^2$). The $Q_{2p}$ values are compared with theoretical predictions of Brown~\cite{brown-ni48}, Ormand~\cite{ormand-ni48} and Cole \cite{cole-ni48}.
The value of the total decay energy, as shown in Table \ref{ni48andfe45-q2pvaluestable}, is in agreement only with the value predicted by Brown.

The angular distribution, built from all available experimental data (current results on top of previous results \cite{miernik-fe45}), is shown in Figure \ref{fig:angulardistributions}(b) and it is compared with the 3-body model calculations for different weights of the $p^2$ orbital, and GCC calculations performed in this work assuming different $V_{pp}$ strengths, represented in orange and purple, respectively, in the same figure. The likelihood probability $L$ and the $\chi^2 \big/dof$ value are shown in Table \ref{tabla-chi2-fe45}. The best agreement is found for the GCC model with the normal strength (100\%) of the nucleon-nucleon force.

The comparison of the angular distribution of $^{45}$Fe with the different theoretical models indicates significant occupancy of the $p^2$ orbital, since the best agreement among the available calculations is obtained for an orbital occupancy $\omega(p^2$)=43\% (3-body model) and confirms the adopted value of the $V_{pp}$ strength (GCC model) for $^{45}$Fe. 
When comparing predicted and experimental partial half-life and $Q_{2p
}$ values, better agreement is found for a similar $p^2$ occupancy for the 3-body model, and the standard  (or even weaker) $V_{pp}$ potential (GCC model). Contrary to the $^{48}$Ni case, the same conclusions are obtained from the different observables - angular distribution, partial half-life and $Q_{2p}$ value - but the observed energy shift in the $Q_{2p}$ values with respect to the values obtained in silicon detector experiments \cite{giovinazzo2002,marek2002,dossat-2005,giovinazzo-2007,miernik-fe45,audirac-2012} needs to be further studied to confirm this conclusion.  

The experimental energy sharing $\frac{E_{p_{i}}}{E_{p_1}+E_{p_2}}$, represented in Figure \ref{fig:energy45Fe}, has been determined with higher precision than in previous TPC experiments \cite{miernik-fe45}, and it is about $1.7$ times narrower than the predicted values in this work (GCC model) and predictions from the 3-body model, represented in purple and orange, respectively, in the same figure. This distribution is not sensitive to the different assumptions of the models, as it can be observed by the overlap of the different predictions.    

The comparison of the energy correlation between the protons gives information mainly from the tunneling process and the Coulomb and centrifugal barriers. Different calculations have been performed varying the total energy of the decay by 100 keV, leading to negligible changes in the width of the distribution. This inconsistency may indicate that $^{45}$Fe is decaying through a different configuration with a larger centrifugal barrier that is partially responsible for this correlation.

\begin{table}[!h]
  \setlength\arrayrulewidth{1.1pt}
 \arrayrulecolor{darkgray}
\centering
\caption{\label{tabla-chi2-fe45} Likelihood probability and $\chi^2 \big/dof$ values obtained when comparing experimental (this work and Miernik {\it et al.}~\cite{miernik-fe45}) and theoretical angular distributions for $^{45}$Fe.
}
 \begin{tabular}{c|c|c}
 \hline
 \hline
  Model & $L(\theta_{pp}$)&$\chi^2 \big/dof$\\  
  \hline 
    GCC $100\% V_{pp}$ & 3.6$\times10^{-13}$&1.31\\
    GCC $125\% V_{pp}$ & 1.4$\times10^{-20}$&4.31\\
    3-body $\omega(p)=10\%$ & 3.7$\times10^{-21}$&16.71\\
    3-body $\omega(p)=24\%$ & 4.4$\times10^{-16}$&3.91\\
    3-body $\omega(p)=43\%$ & 2.3$\times10^{-15}$&2.59\\
    \hline
    \hline
\end{tabular}
\end{table}

 \section{Conclusions}

Observables of two-proton radioactivity from nuclear ground states have been measured for $^{48}$Ni and $^{45}$Fe during an experiment in 2021 at GANIL. New GCC calculations, taking into account the three main features of this rare decay mode - nuclear structure, asymptotic region and decay dynamics - have been performed aiming at a deeper understanding of two-proton radioactivity.  

The comparison of the different two-proton radioactivity  observables (angular distribution, half-life, total decay energy, sharing of energy) between experiment and theory should provide a unique interpretation of the two-proton decay characteristics. The available experimental results for $^{48}$Ni are consistent with each other. The comparisons of the angular distribution and the half-life values with theory lead to opposite conclusions for both theoretical models. The experimental angular distribution fits better to GCC calculations performed using a $V_{pp}$=100\% value and 3-body model calculations using a low p occupancy $\omega(p)$=2\%. Inversely, when comparing half-life values and total energy of the decay, a better agreement is found for $V_{pp}$=125\% and much higher p occupancy values for GCC and 3-body model, respectively.
A contribution of the $s$-wave continuum, highly suppressed in the present GCC calculations, 
could possibly retrieve the agreement with the experimental results. When comparing these observables for $^{45}$Fe for both approaches, compatible conclusions are extracted, but here a lack of experimental agreement is found between the results of the present work and previously reported values for half-life and total decay energy. A possible experimental bias for the $Q_{2p}$ values between TPC-type experiments and silicon-detector based measurements has to be investigated. An experiment using both (ACTAR TPC and Si detectors) at once in tandem mode could be performed at LISE3 for that.

The sharing of the energy of the protons has been obtained with higher precision with respect to previous work \cite{miernik-fe45}, and it has been found to be much narrower than the predicted values from theory. This indicates that $^{45}$Fe may decay through a different configuration with higher angular momentum components. However, before reaching such a conclusion, the origin of experimental inconsistencies found in this work has to be clarified.

These results show the complexity of the two-proton radioactivity process and prompt the need for further experimental and theoretical efforts to disentangle these discrepancies. This may then lead to a consistent description of the process, which could shed light on different open questions in nuclear physics like quantum tunneling, pairing effect in nuclear medium and continuum, as well as fermionic dynamics in open quantum systems. 

\section{Acknowledgement}
We are grateful to the ion-source and accelerator staff at the Grand Accélérateur National d’Ions Lourds (GANIL) for the provision of a stable, high-intensity, primary $^{58}$Ni beam. This work was supported by European Union’s Horizon 2020 Framework research and innovation programme 654002 (ENSAR2). The ACTAR TPC development was funded by the European Research Council under the European Union’s Seventh Framework Program (FP7/2007-2013)/ERC grant agreement no 335593 and by the Conseil Régional d’Aquitaine, France (grant no 2014-1R60402 - 00003319). 
A. Ortega Moral wants to acknowledge A. Navin for helpful discussions and comments during the drafting of this paper.


\section*{References}
\begin{thebibliography}{36}%
\makeatletter
\providecommand \@ifxundefined [1]{%
 \@ifx{#1\undefined}
}%
\providecommand \@ifnum [1]{%
 \ifnum #1\expandafter \@firstoftwo
 \else \expandafter \@secondoftwo
 \fi
}%
\providecommand \@ifx [1]{%
 \ifx #1\expandafter \@firstoftwo
 \else \expandafter \@secondoftwo
 \fi
}%
\providecommand \natexlab [1]{#1}%
\providecommand \enquote  [1]{``#1''}%
\providecommand \bibnamefont  [1]{#1}%
\providecommand \bibfnamefont [1]{#1}%
\providecommand \citenamefont [1]{#1}%
\providecommand \href@noop [0]{\@secondoftwo}%
\providecommand \href [0]{\begingroup \@sanitize@url \@href}%
\providecommand \@href[1]{\@@startlink{#1}\@@href}%
\providecommand \@@href[1]{\endgroup#1\@@endlink}%
\providecommand \@sanitize@url [0]{\catcode `\\12\catcode `\$12\catcode
  `\&12\catcode `\#12\catcode `\^12\catcode `\_12\catcode `\%12\relax}%
\providecommand \@@startlink[1]{}%
\providecommand \@@endlink[0]{}%
\providecommand \url  [0]{\begingroup\@sanitize@url \@url }%
\providecommand \@url [1]{\endgroup\@href {#1}{\urlprefix }}%
\providecommand \urlprefix  [0]{URL }%
\providecommand \Eprint [0]{\href }%
\providecommand \doibase [0]{http://dx.doi.org/}%
\providecommand \selectlanguage [0]{\@gobble}%
\providecommand \bibinfo  [0]{\@secondoftwo}%
\providecommand \bibfield  [0]{\@secondoftwo}%
\providecommand \translation [1]{[#1]}%
\providecommand \BibitemOpen [0]{}%
\providecommand \bibitemStop [0]{}%
\providecommand \bibitemNoStop [0]{.\EOS\space}%
\providecommand \EOS [0]{\spacefactor3000\relax}%
\providecommand \BibitemShut  [1]{\csname bibitem#1\endcsname}%
\let\auto@bib@innerbib\@empty
\bibitem [{\citenamefont {{V. Goldanskii}}(1960)}]{goldanskii-pemission}%
  \BibitemOpen
  \bibfield  {author} {\bibinfo {author} {\bibnamefont {{V. Goldanskii}}},\
  }\href@noop {} {\bibfield  {journal} {\bibinfo  {journal} {Nucl. Phys.}\
  }\textbf {\bibinfo {volume} {19}},\ \bibinfo {pages} {482} (\bibinfo {year}
  {1960})}\BibitemShut {NoStop}%
\bibitem [{\citenamefont {{B.A. Brown {\it et al.}}}(2019)}]{brown-hybrid2019}%
  \BibitemOpen
  \bibfield  {author} {\bibinfo {author} {\bibnamefont {{B.A. Brown {\it et
  al.}}}},\ }\href@noop {} {\bibfield  {journal} {\bibinfo  {journal} {Phys.
  Rev. C}\ }\textbf {\bibinfo {volume} {100}},\ \bibinfo {pages} {054332}
  (\bibinfo {year} {2019})}\BibitemShut {NoStop}%
\bibitem [{\citenamefont {{K. Miernik {\it et
  al.}}}(2007{\natexlab{a}})}]{miernik-fe45}%
  \BibitemOpen
  \bibfield  {author} {\bibinfo {author} {\bibnamefont {{K. Miernik {\it et
  al.}}}},\ }\href@noop {} {\bibfield  {journal} {\bibinfo  {journal} {Phys.
  Rev. Lett.}\ }\textbf {\bibinfo {volume} {99}},\ \bibinfo {pages} {192501}
  (\bibinfo {year} {2007}{\natexlab{a}})}\BibitemShut {NoStop}%
\bibitem [{\citenamefont {{T. Goigoux {\it et al.}}}(2016)}]{goigoux2016}%
  \BibitemOpen
  \bibfield  {author} {\bibinfo {author} {\bibnamefont {{T. Goigoux {\it et
  al.}}}},\ }\href@noop {} {\bibfield  {journal} {\bibinfo  {journal} {Phys.
  Rev. Lett.}\ }\textbf {\bibinfo {volume} {117}},\ \bibinfo {pages} {162501}
  (\bibinfo {year} {2016})}\BibitemShut {NoStop}%
\bibitem [{\citenamefont {{L. V. Grigorenko {\it et
  al.}}}(2000)}]{grigorenko-2000-threebody}%
  \BibitemOpen
  \bibfield  {author} {\bibinfo {author} {\bibnamefont {{L. V. Grigorenko {\it
  et al.}}}},\ }\href@noop {} {\bibfield  {journal} {\bibinfo  {journal} {Phys.
  Rev. Lett.}\ }\textbf {\bibinfo {volume} {85}},\ \bibinfo {pages} {22}
  (\bibinfo {year} {2000})}\BibitemShut {NoStop}%
\bibitem [{\citenamefont {{S. M. Wang, W. Nazarewicz}}(2018)}]{nazarewicz2018}%
  \BibitemOpen
  \bibfield  {author} {\bibinfo {author} {\bibnamefont {{S. M. Wang, W.
  Nazarewicz}}},\ }\href@noop {} {\bibfield  {journal} {\bibinfo  {journal}
  {Phys. Rev. C}\ }\textbf {\bibinfo {volume} {120}},\ \bibinfo {pages}
  {212502} (\bibinfo {year} {2018})}\BibitemShut {NoStop}%
\bibitem [{\citenamefont {{J. Giovinazzo {\it et
  al.}}}(2002)}]{giovinazzo2002}%
  \BibitemOpen
  \bibfield  {author} {\bibinfo {author} {\bibnamefont {{J. Giovinazzo {\it et
  al.}}}},\ }\href@noop {} {\bibfield  {journal} {\bibinfo  {journal} {Phys.
  Rev. Lett.}\ }\textbf {\bibinfo {volume} {89}},\ \bibinfo {pages} {102501}
  (\bibinfo {year} {2002})}\BibitemShut {NoStop}%
\bibitem [{\citenamefont {{M. Pf\"utzner {\it et al.}}}(2002)}]{marek2002}%
  \BibitemOpen
  \bibfield  {author} {\bibinfo {author} {\bibnamefont {{M. Pf\"utzner {\it et
  al.}}}},\ }\href@noop {} {\bibfield  {journal} {\bibinfo  {journal} {Eur.
  Phys. J. A}\ }\textbf {\bibinfo {volume} {14}},\ \bibinfo {pages} {279}
  (\bibinfo {year} {2002})}\BibitemShut {NoStop}%
\bibitem [{\citenamefont {{B.~Blank {\it et al.}}}(2008)}]{cenbg-tpc}%
  \BibitemOpen
  \bibfield  {author} {\bibinfo {author} {\bibnamefont {{B.~Blank {\it et
  al.}}}},\ }\href@noop {} {\bibfield  {journal} {\bibinfo  {journal} {Nucl.
  Instrum. Meth. Phys,}\ }\textbf {\bibinfo {volume} {B266}},\ \bibinfo {pages}
  {4606} (\bibinfo {year} {2008})}\BibitemShut {NoStop}%
\bibitem [{\citenamefont {{K. Miernik {\it et
  al.}}}(2007{\natexlab{b}})}]{otpc2007}%
  \BibitemOpen
  \bibfield  {author} {\bibinfo {author} {\bibnamefont {{K. Miernik {\it et
  al.}}}},\ }\href@noop {} {\bibfield  {journal} {\bibinfo  {journal} {Nucl.
  Instrum. Meth. Phys. Res. A}\ }\textbf {\bibinfo {volume} {581}},\ \bibinfo
  {pages} {194} (\bibinfo {year} {2007}{\natexlab{b}})}\BibitemShut {NoStop}%
\bibitem [{\citenamefont {{T.~Roger {\it et al.}}}(2018)}]{roger2018}%
  \BibitemOpen
  \bibfield  {author} {\bibinfo {author} {\bibnamefont {{T.~Roger {\it et
  al.}}}},\ }\href@noop {} {\bibfield  {journal} {\bibinfo  {journal} {Nucl.
  Instrum. Meth. Phys. Res.}\ }\textbf {\bibinfo {volume} {A 895}},\ \bibinfo
  {pages} {126} (\bibinfo {year} {2018})}\BibitemShut {NoStop}%
\bibitem [{\citenamefont {{B.~Mauss {\it et al.}}}(2019)}]{mauss2019}%
  \BibitemOpen
  \bibfield  {author} {\bibinfo {author} {\bibnamefont {{B.~Mauss {\it et
  al.}}}},\ }\href@noop {} {\bibfield  {journal} {\bibinfo  {journal} {Nucl.
  Instrum. Meth. Phys. Res.}\ }\textbf {\bibinfo {volume} {A 940}},\ \bibinfo
  {pages} {498} (\bibinfo {year} {2019})}\BibitemShut {NoStop}%
\bibitem [{\citenamefont {{L. V. Grigorenko {\it et
  al.}}}(2017)}]{grigorenko-2017}%
  \BibitemOpen
  \bibfield  {author} {\bibinfo {author} {\bibnamefont {{L. V. Grigorenko {\it
  et al.}}}},\ }\href@noop {} {\bibfield  {journal} {\bibinfo  {journal} {Phys.
  Rev. C}\ }\textbf {\bibinfo {volume} {95}},\ \bibinfo {pages} {021601}
  (\bibinfo {year} {2017})}\BibitemShut {NoStop}%
\bibitem [{\citenamefont {{S. M. Wang and W. Nazarewicz}}(2021)}]{wangPRL2021}%
  \BibitemOpen
  \bibfield  {author} {\bibinfo {author} {\bibnamefont {{S. M. Wang and W.
  Nazarewicz}}},\ }\href@noop {} {\bibfield  {journal} {\bibinfo  {journal}
  {Phys. Rev. Lett.}\ }\textbf {\bibinfo {volume} {126}},\ \bibinfo {pages}
  {142501} (\bibinfo {year} {2021})}\BibitemShut {NoStop}%
\bibitem [{\citenamefont {{S. M. Wang {\it et al.}}}(2022)}]{wangjpg2022}%
  \BibitemOpen
  \bibfield  {author} {\bibinfo {author} {\bibnamefont {{S. M. Wang {\it et
  al.}}}},\ }\href@noop {} {\bibfield  {journal} {\bibinfo  {journal} {J. Phys.
  G}\ }\textbf {\bibinfo {volume} {49}},\ \bibinfo {pages} {10LT02} (\bibinfo
  {year} {2022})}\BibitemShut {NoStop}%
\bibitem [{\citenamefont {{S. M. Wang {\it et al.}}}(2017)}]{wangprc2017}%
  \BibitemOpen
  \bibfield  {author} {\bibinfo {author} {\bibnamefont {{S. M. Wang {\it et
  al.}}}},\ }\href@noop {} {\bibfield  {journal} {\bibinfo  {journal} {Phys.
  Rev. C}\ }\textbf {\bibinfo {volume} {96}},\ \bibinfo {pages} {044307}
  (\bibinfo {year} {2017})}\BibitemShut {NoStop}%
\bibitem [{\citenamefont {{S. M. Wang and W. Nazarewicz}}(2018)}]{wangPRL2018}%
  \BibitemOpen
  \bibfield  {author} {\bibinfo {author} {\bibnamefont {{S. M. Wang and W.
  Nazarewicz}}},\ }\href@noop {} {\bibfield  {journal} {\bibinfo  {journal}
  {Phys. Rev. Lett.}\ }\textbf {\bibinfo {volume} {120}},\ \bibinfo {pages}
  {212502} (\bibinfo {year} {2018})}\BibitemShut {NoStop}%
\bibitem [{\citenamefont {{S. M. Wang and W.
  Nazarewicz}}(2024)}]{wanginpreparation}%
  \BibitemOpen
  \bibfield  {author} {\bibinfo {author} {\bibnamefont {{S. M. Wang and W.
  Nazarewicz}}},\ }\href@noop {} {\bibfield  {journal} {\bibinfo  {journal}
  {Correlations for mid-heavy two-proton emitters, in preparation}\ } (\bibinfo
  {year} {2024})}\BibitemShut {NoStop}%
\bibitem [{\citenamefont {{H. Furutani {\it et al.}}}(1979)}]{Furutani1979}%
  \BibitemOpen
  \bibfield  {author} {\bibinfo {author} {\bibnamefont {{H. Furutani {\it et
  al.}}}},\ }\href@noop {} {\bibfield  {journal} {\bibinfo  {journal} {Prog.
  Theor. Phys.}\ }\textbf {\bibinfo {volume} {62}},\ \bibinfo {pages} {981}
  (\bibinfo {year} {1979})}\BibitemShut {NoStop}%
\bibitem [{\citenamefont {{S. Cwiok {\it et al.}}}(1987)}]{Cwiok1987}%
  \BibitemOpen
  \bibfield  {author} {\bibinfo {author} {\bibnamefont {{S. Cwiok {\it et
  al.}}}},\ }\href@noop {} {\bibfield  {journal} {\bibinfo  {journal} {Comput.
  Phys. Comm.}\ }\textbf {\bibinfo {volume} {46}},\ \bibinfo {pages} {379}
  (\bibinfo {year} {1987})}\BibitemShut {NoStop}%
\bibitem [{\citenamefont {{A.C. Mueller and R. Anne}}(1991)}]{anne1994}%
  \BibitemOpen
  \bibfield  {author} {\bibinfo {author} {\bibnamefont {{A.C. Mueller and R.
  Anne}}},\ }\href@noop {} {\bibfield  {journal} {\bibinfo  {journal} {Nucl.
  Instrum. Meth. Phys. Res.}\ }\textbf {\bibinfo {volume} {B56}},\ \bibinfo
  {pages} {559} (\bibinfo {year} {1991})}\BibitemShut {NoStop}%
\bibitem [{\citenamefont {{S. Ottini {\it et al.}}}(1999)}]{cats1999}%
  \BibitemOpen
  \bibfield  {author} {\bibinfo {author} {\bibnamefont {{S. Ottini {\it et
  al.}}}},\ }\href@noop {} {\bibfield  {journal} {\bibinfo  {journal} {Nucl.
  Instrum. Meth. Phys. Res. A}\ }\textbf {\bibinfo {volume} {431}},\ \bibinfo
  {pages} {476} (\bibinfo {year} {1999})}\BibitemShut {NoStop}%
\bibitem [{\citenamefont {{C. Dossat {\it et al.}}}(2005)}]{dossat-2005}%
  \BibitemOpen
  \bibfield  {author} {\bibinfo {author} {\bibnamefont {{C. Dossat {\it et
  al.}}}},\ }\href@noop {} {\bibfield  {journal} {\bibinfo  {journal} {Phys.
  Rev. C.}\ }\textbf {\bibinfo {volume} {72}},\ \bibinfo {pages} {054315}
  (\bibinfo {year} {2005})}\BibitemShut {NoStop}%
\bibitem [{\citenamefont {{E.~Pollacco {\it et al.}}}(2018)}]{pollacco2018}%
  \BibitemOpen
  \bibfield  {author} {\bibinfo {author} {\bibnamefont {{E.~Pollacco {\it et
  al.}}}},\ }\href@noop {} {\bibfield  {journal} {\bibinfo  {journal} {Nucl.
  Instrum. Meth. Phys. Res.}\ }\textbf {\bibinfo {volume} {A887}},\ \bibinfo
  {pages} {81} (\bibinfo {year} {2018})}\BibitemShut {NoStop}%
\bibitem [{\citenamefont {{J. Giovinazzo {\it et al.}}}(2022)}]{jerome2022}%
  \BibitemOpen
  \bibfield  {author} {\bibinfo {author} {\bibnamefont {{J. Giovinazzo {\it et
  al.}}}},\ }\href@noop {} {\bibfield  {journal} {\bibinfo  {journal} {Nucl.
  Instrum. Meth. Phys. Res. A}\ }\textbf {\bibinfo {volume} {1042}},\ \bibinfo
  {pages} {167447} (\bibinfo {year} {2022})}\BibitemShut {NoStop}%
\bibitem [{\citenamefont {{T. Ziegler {\it et al.}}}()}]{SRIM}%
  \BibitemOpen
  \bibfield  {author} {\bibinfo {author} {\bibnamefont {{T. Ziegler {\it et
  al.}}}},\ }\href@noop {} {\bibinfo  {journal} {http://www.srim.org}\
  }\BibitemShut {NoStop}%
\bibitem [{\citenamefont {{M. Pomorski {\it et al.}}}(2014)}]{Pomorski-ni48}%
  \BibitemOpen
\bibfield  {journal} {  }\bibfield  {author} {\bibinfo {author} {\bibnamefont
  {{M. Pomorski {\it et al.}}}},\ }\href@noop {} {\bibfield  {journal}
  {\bibinfo  {journal} {Phys. Rev.}\ }\textbf {\bibinfo {volume} {C90}},\
  \bibinfo {pages} {014311} (\bibinfo {year} {2014})}\BibitemShut {NoStop}%
\bibitem [{\citenamefont {{K.-H. Schmidt}}(1984)}]{pocascuentas}%
  \BibitemOpen
  \bibfield  {author} {\bibinfo {author} {\bibnamefont {{K.-H. Schmidt}}},\
  }\href@noop {} {\bibfield  {journal} {\bibinfo  {journal} {Zeitschrift f\"ur
  Physik A, Atoms and Nuclei}\ }\textbf {\bibinfo {volume} {316}},\ \bibinfo
  {pages} {19} (\bibinfo {year} {1984})}\BibitemShut {NoStop}%
\bibitem [{\citenamefont {{B.A. Brown}}(1991)}]{brown-ni48}%
  \BibitemOpen
  \bibfield  {author} {\bibinfo {author} {\bibnamefont {{B.A. Brown}}},\
  }\href@noop {} {\bibfield  {journal} {\bibinfo  {journal} {Phys. Rev. C}\
  }\textbf {\bibinfo {volume} {43}},\ \bibinfo {pages} {R1513} (\bibinfo {year}
  {1991})}\BibitemShut {NoStop}%
\bibitem [{\citenamefont {{W.E. Ormand}}(1997)}]{ormand-ni48}%
  \BibitemOpen
  \bibfield  {author} {\bibinfo {author} {\bibnamefont {{W.E. Ormand}}},\
  }\href@noop {} {\bibfield  {journal} {\bibinfo  {journal} {Phys. Rev. C}\
  }\textbf {\bibinfo {volume} {55}},\ \bibinfo {pages} {2407} (\bibinfo {year}
  {1997})}\BibitemShut {NoStop}%
\bibitem [{\citenamefont {{B.J. Cole}}(1996)}]{cole-ni48}%
  \BibitemOpen
  \bibfield  {author} {\bibinfo {author} {\bibnamefont {{B.J. Cole}}},\
  }\href@noop {} {\bibfield  {journal} {\bibinfo  {journal} {Phys. Rev. C}\
  }\textbf {\bibinfo {volume} {54}},\ \bibinfo {pages} {1240} (\bibinfo {year}
  {1996})}\BibitemShut {NoStop}%
\bibitem [{\citenamefont {{D.R. Thomson {\it et al.}}}(1977)}]{minnesotaforce}%
  \BibitemOpen
  \bibfield  {author} {\bibinfo {author} {\bibnamefont {{D.R. Thomson {\it et
  al.}}}},\ }\href@noop {} {\bibfield  {journal} {\bibinfo  {journal} {Nucl.
  Phys. A}\ }\textbf {\bibinfo {volume} {286}},\ \bibinfo {pages} {53}
  (\bibinfo {year} {1977})}\BibitemShut {NoStop}%
\bibitem [{\citenamefont {{L. Audirac {\it et al.}}}(2012)}]{audirac-2012}%
  \BibitemOpen
  \bibfield  {author} {\bibinfo {author} {\bibnamefont {{L. Audirac {\it et
  al.}}}},\ }\href@noop {} {\bibfield  {journal} {\bibinfo  {journal} {Eur.
  Phys. J. A}\ }\textbf {\bibinfo {volume} {99}},\ \bibinfo {pages} {12179}
  (\bibinfo {year} {2012})}\BibitemShut {NoStop}%
\bibitem [{Note1()}]{Note1}%
  \BibitemOpen
  \bibinfo {note} {The same analysis was also performed for $^{49}$Ni, but no
  two-proton events were found in this case.}\BibitemShut {Stop}%
\bibitem [{\citenamefont {{M. Wang, W. Nazarewicz}}(2023)}]{gcc-2023}%
  \BibitemOpen
  \bibfield  {author} {\bibinfo {author} {\bibnamefont {{M. Wang, W.
  Nazarewicz}}},\ }\href@noop {} {\bibfield  {journal} {\bibinfo  {journal}
  {private comunication}\ } (\bibinfo {year} {2023})}\BibitemShut {NoStop}%
\bibitem [{\citenamefont {{J. Giovinazzo {\it et
  al.}}}(2007)}]{giovinazzo-2007}%
  \BibitemOpen
  \bibfield  {author} {\bibinfo {author} {\bibnamefont {{J. Giovinazzo {\it et
  al.}}}},\ }\href@noop {} {\bibfield  {journal} {\bibinfo  {journal} {Phys.
  Rev. Lett.}\ }\textbf {\bibinfo {volume} {99}},\ \bibinfo {pages} {102501}
  (\bibinfo {year} {2007})}\BibitemShut {NoStop}%
\end{thebibliography}
\end{document}